\documentstyle[prl,aps,floats,amsmath,epsf,graphicx]{revtex}

\begin{document}
\preprint{astro-ph/0009131}
\draft

\newcommand\eq[1]{Eq.~(\ref{#1})}
\newcommand\eqs[2]{Eqs.~(\ref{#1}) and (\ref{#2})}
\newcommand\eqss[3]{Eqs.~(\ref{#1}), (\ref{#2}) and (\ref{#3})}
\newcommand\eqsss[4]{Eqs.~(\ref{#1}), (\ref{#2}), (\ref{#3})
and (\ref{#4})}
\newcommand\eqssss[5]{Eqs.~(\ref{#1}), (\ref{#2}), (\ref{#3}),
(\ref{#4}) and (\ref{#5})}
\newcommand\eqst[2]{Eqs.~(\ref{#1})--(\ref{#2})}

\newcommand\ee{\end{equation}}
\newcommand\be{\begin{equation}}
\newcommand\eea{\end{eqnarray}}
\newcommand\bea{\begin{eqnarray}}
\renewcommand{\topfraction}{0.99}

\twocolumn[\hsize\textwidth\columnwidth\hsize\csname
@twocolumnfalse\endcsname

\title{Adiabatic and entropy perturbations from inflation}

\author{Christopher Gordon, David Wands, Bruce A.~Bassett and
Roy Maartens}

\address{Relativity and Cosmology Group,
School of Computer Science and Mathematics, University of
Portsmouth, Portsmouth~PO1~2EG, Britain}


\maketitle

\begin{abstract}

We study adiabatic (curvature) and entropy (isocurvature)
perturbations produced during a period of cosmological inflation that
is driven by multiple scalar fields with an arbitrary interaction
potential.  A local rotation in field space is performed to separate
out the adiabatic and entropy modes.  The resulting field equations
show explicitly how on large scales entropy perturbations can source
adiabatic perturbations if the background solution follows a curved
trajectory in field space, and how adiabatic perturbations cannot
source entropy perturbations in the long-wavelength limit. It is the
effective mass of the entropy field that determines the amplitude of
entropy perturbations during inflation. We present two applications of
the equations. First, we show why one in general expects the adiabatic
and entropy perturbations to be correlated at the end of inflation,
and calculate the cross-correlation in the context of a double
inflation model with two non-interacting fields. Second, we consider
two-field preheating after inflation, examining conditions under
which entropy perturbations can alter the large-scale curvature
perturbation and showing how our new formalism has advantages in
numerical stability when the background solution follows a non-trivial
trajectory in field space.

\end{abstract}

\pacs{PACS numbers: 98.80.Cq \hfill Accepted for publication in Phys.\
  Rev.\ D. \hfill astro-ph/0009131}

\vskip2pc]

\section{Introduction}

A period of accelerated expansion -- inflation -- in the early
universe has become the standard model for the origin of structure
in the universe. Inhomogeneities in the present matter
distribution can be traced back to quantum fluctuations in the
fields driving inflation which are stretched beyond the Hubble
scale during inflation. In the simplest models of inflation driven by a
single scalar field, these fluctuations produce a primordial
adiabatic spectrum
whose amplitude can be characterized by the comoving curvature
perturbation ${\cal R}$, which remains constant on super-Hubble
scales until the perturbation comes back within the Hubble scale
long after inflation has ended.

As soon as one considers more than one scalar field, one must also
consider the role of non-adiabatic fluctuations. This can have
important consequences, both in affecting the evolution of the
curvature perturbation (often referred to as the `adiabatic
perturbation'), but also in the possibility of seeding isocurvature
(or `entropy') perturbations after inflation.

Previous studies have demonstrated that non-adiabatic pressure
perturbations can alter the curvature perturbation on super-Hubble
scales either during inflation~\cite{GBW,Shinji} or
after~\cite{HK,Hu,WMLL}.  A general formalism to evaluate the
curvature perturbation at the end of inflation in multiple field
models was developed in Ref.~\cite{SS}.  In the presence of
non-adiabatic fluctuations, one must follow the evolution of perturbed
fields on super-Hubble scales, in particular tracking the perturbation
in the integrated expansion~\cite{Star,SS,Salopek,SasTan,LR,WMLL}, in
order to 
evaluate the large-scale curvature perturbation at late
times~\cite{Polarski:1992dq,PolSta94,SY,GBW,SS,Salopek,SasTan,MukSte,julien}.

However no similar formalism has been developed so far to evaluate
the isocurvature perturbation in the general case. Instead,
isocurvature perturbations have been studied in a number of
particular models of inflation~\cite{many,Polarski:1992dq,PolSta94}.
These fluctuations typically arise as baryon modes
(e.g.~\cite{baryon}) or cold dark matter modes~\cite{P}, but
neutrino isocurvature modes have also been considered~\cite{BMT}.
Recently, it has been pointed out~\cite{Langlois,BMT2} that it is
rather natural to expect the curvature and isocurvature
perturbations to be correlated, which yields distinctive
observational results~\cite{Langlois2}, in contrast to the
isocurvature perturbations usually tested against
observations~\cite{isocmb}.

In this paper we will develop a general formalism to study the
evolution of both curvature and isocurvature perturbations in a wide
class of multi-field inflation models by decomposing field
perturbations into perturbations along the background trajectory in
field space (the adiabatic field perturbation), and orthogonal to the
background trajectory (the entropy field).  We allow an arbitrary
interaction potential for the fields, and, although we concentrate
upon the case of two scalar fields, the general approach can be easily
extended to $N$ fields, where there will be $N-1$ entropy fields
orthogonal to the background trajectory. This was done for a specific
assisted inflation model in Ref.~\cite{Karim}. We will work in the
metric based approach of Bardeen~\cite{Bardeen} in order to define
gauge-invariant cosmological perturbations, but our formalism can also
be applied to the study of multiple scalar fields in other
approaches~\cite{Hwang,Ellis,Durrer}.

We begin by reviewing the standard results obtained in single field
models, emphasizing the suppression of non-adiabatic fluctuations on
large-scales. We then extend our analysis to general two-field models,
defining an adiabatic field and an entropy field, whose fluctuations,
though uncorrelated on small scales, may develop correlations through
the subsequent evolution. We present two specific models of two-field
inflation, one with non-interacting fields, the other a model of
interacting fields which undergo preheating after inflation.

\section{Perturbation equations for multiple scalar fields}

We consider $N$ scalar fields  with  Lagrangian density:
\begin{gather}
{\cal L} = -V(\varphi_1,\cdots,\varphi_N)
 -\frac{1}{2} \sum_{I=1}^{N} g^{\mu\nu} \varphi_{I,\mu}\varphi_{I,\nu}
 \,,
\label{lag}
\end{gather}
and minimal coupling to gravity. 
In order to study the evolution of linear perturbations in the scalar
fields, we make the standard splitting $\varphi_I(t,{\bf x})
\to\varphi_I(t)+\delta\varphi_I(t,{\bf x})$. 
The field equations, derived from Eq.~(\ref{lag}) for the background
homogeneous fields, are
\begin{equation}
\label{eq:KG} 
\ddot{\varphi}_I + 3H\dot{\varphi}_I + V_{\varphi_I} = 0\,,
\end{equation}
where $V_{x} = {\partial V}/{\partial x}$, and the Hubble rate, $H$,
in a spatially flat Friedmann-Robertson-Walker (FRW) universe, is
determined by the Friedman equation:
\begin{gather}
H^2 = \left(\frac{\dot{a}}{a}\right)^2 =\frac{8\pi G}{3} \left[
V(\varphi_I) + \frac{1}{2}  \sum_I \dot\varphi_I^{~2}
 \right]\,, \label{eq:hubble}
\end{gather}
with $a(t)$ the FRW scale factor.

Consistent study of the linear field fluctuations $\delta\varphi_I$
requires that we also consider linear scalar perturbations of the
metric, corresponding to the line element\footnote{
We follow the notation of Ref.~\cite{MFB}, apart from our use of
$A$ rather than $\phi$ as the perturbation in the lapse function.
}
\begin{eqnarray}
ds^2 &=& - (1+2A)dt^2 + 2aB_{,i}dx^idt \nonumber\\ &&~{} +
a^2\left[ (1-2\psi)\delta_{ij} + 2E_{,ij}\right] dx^idx^j \,,
\end{eqnarray}
where we have not at this stage specified any particular choice of
gauge~\cite{MFB,Bardeen,KS}.

Scalar field perturbations, with comoving wavenumber $k=2\pi
a/\lambda$ for a mode with physical wavelength $\lambda$, then
obey the perturbation equations
\begin{eqnarray}
&& \ddot{\delta\varphi}_I + 3H\dot{\delta\varphi}_I
 + \frac{k^2}{a^2} \delta\varphi_I + \sum_J V_{\varphi_I\varphi_J}
\delta\varphi_J
  \nonumber\\ &&~~{}=
-2V_{\varphi_I}A + \dot\varphi_I \left[ \dot{A} + 3\dot{\psi} +
\frac{k^2}{a^2} (a^2\dot{E}-aB) \right] \,. \label{eq:perturbation}
\end{eqnarray}
The metric terms on the right-hand-side, induced  by the scalar
field perturbations, obey the energy and momentum constraints
\begin{eqnarray}
3H\left(\dot\psi+HA\right) + 
\frac{k^2}{a^2}\left[\psi+H(a^2\dot{E}-aB)\right] &=& -4\pi G \delta\rho \,,
\label{eq:densitycon}
\\
\dot\psi + HA &=& -4\pi G \delta q \,.
\label{eq:mtmcon}
\end{eqnarray}
The total energy and momentum perturbations are given in terms of
the scalar field perturbations by
\begin{eqnarray}
\delta\rho &=& \sum_I\left[
 \dot\varphi_I \left( \dot{\delta\varphi}_I -\dot\varphi_I A \right)
 + V_{\varphi_I}\delta\varphi_I \right]
\label{eq:density} \\ 
\delta q_{,i} &=& - \sum_I \dot{\varphi}_I
\delta\varphi_{I,i} \,. 
\label{eq:mtm}
\end{eqnarray}
These two equations can be combined to construct a gauge-invariant
quantity, the comoving density perturbation~\cite{Bardeen}
\begin{eqnarray}
\label{def:epsilonm} \epsilon_m &\equiv& \delta\rho -3H\delta q
\nonumber\\ &=& \sum_I \left[ \dot\varphi_I \left(
\dot{\delta\varphi}_I - \dot\varphi_I A \right) - \ddot\varphi_I
\delta\varphi_I \right]\,,
\end{eqnarray}
which is sometimes used to represent the total matter perturbation.

Because the anisotropic stress vanishes to linear order for scalar
fields minimally coupled to gravity, we have a further constraint on
the metric perturbations:
\begin{equation}
\label{eq:aniso} \left( a^2\dot{E}-aB
\right)^{\displaystyle{\cdot}} + H \left( a^2\dot{E}-aB \right) +
\psi - A = 0 \,.
\end{equation}

The coupled perturbation
equations~(\ref{eq:perturbation})--(\ref{eq:mtm}) and
(\ref{eq:aniso}) are probably most often solved in the zero-shear
(or longitudinal or conformal Newtonian) gauge, in which
$a^2\dot{E}_\ell-aB_\ell=0$~\cite{MFB}. The two remaining metric
perturbation variables which appear in the scalar field
perturbation equation, $A_\ell=\Phi$ and $\psi_\ell=\Psi$, are
then equal in the absence of any anisotropic stress by
Eq.~(\ref{eq:aniso}).

Another useful choice is the spatially flat gauge, in which
$\psi_Q=0$~\cite{KS,Hwang}.  The scalar field perturbations in this
gauge are sometimes referred to as the Sasaki or Mukhanov
variables~\cite{SM}, which have the gauge-invariant definition
\begin{equation}
\label{def:QI}
Q_{I} \equiv \delta\varphi_I + \frac{\dot\varphi_I}{H}\psi \,.
\end{equation}
The shear perturbation in the spatially flat gauge is simply
related to the curvature perturbation, $\Psi$, in the zero-shear
gauge:
\begin{equation}
a^2\dot{E}_Q - aB_Q = a^2\dot{E} - aB + \frac{1}{H}\psi = \frac{1}{H}\Psi \,.
\end{equation}
The energy and momentum constraints, Eqs.~(\ref{eq:densitycon})
and~(\ref{eq:mtmcon}), in the spatially flat gauge thus yield
\begin{eqnarray}
\label{eq:Psi}
\frac{k^2}{a^2} \Psi
 &=& - {4\pi G} \epsilon_m
 \,,\\
\label{eq:AQ} HA_Q &=& - 4\pi G \delta q_Q \,,
\end{eqnarray}
where $\epsilon_m$ is given in Eq.~(\ref{def:epsilonm}), and
from Eq.~(\ref{eq:mtm}) we have $\delta q_Q = -\sum_I\dot\varphi_I
Q_I$. 

The equations of motion, Eq.~(\ref{eq:perturbation}), rewritten in
terms of the Sasaki-Mukhanov variables, and using
Eqs.~(\ref{eq:Psi}) and (\ref{eq:AQ}) to eliminate the metric
perturbation terms in the spatially flat gauge, become~\cite{TN}:
\begin{eqnarray}
&&\ddot Q_I + 3 H \dot Q_I + \frac{k^2}{a^2} Q_I  \nonumber\\
&&~~{}
 + \sum_{J} \left[ V_{\varphi_I\varphi_J} - \frac{8 \pi G}{a^3}
\left( \frac{a^3}{H} \dot \varphi_I \dot \varphi_J
\right)^{\displaystyle{\cdot}} \right] Q_J = 0 \,. \label{eq:multi
Q}
\end{eqnarray}

\subsection{Curvature and  entropy perturbations}

The comoving curvature perturbation~\cite{Lukash,Lyth85} is given by
\begin{eqnarray}
\label{def:calR} {\cal R} &\equiv& \psi - \frac{H}{\rho+p} \delta q
 \nonumber\\
&=&  \sum_I \left(\frac{\dot\varphi_I}{\sum_J \dot\varphi_J^2}
\right) Q_{I} \,.
\end{eqnarray}
This can also be given in terms of the metric perturbations in the
longitudinal gauge as~\cite{MFB}
\begin{equation}
{\cal R} = \Psi - {H\over\dot{H}} \left( \dot\Psi+H\Phi \right) \,.
\end{equation}

For comparison we give the curvature perturbation on
uniform-density hypersurfaces,
\begin{equation}
{}-\zeta \equiv \psi + H {\delta\rho\over \dot\rho} \,,
\end{equation}
first introduced by Bardeen, Steinhardt and Turner~\cite{BST} as a
conserved quantity for adiabatic perturbations on large
scales~\cite{MarSch,WMLL}. It is related to the comoving curvature
perturbation in Eq.~(\ref{def:calR}) by a gauge transformation
\begin{equation}
{}-\zeta = {\cal R} + {2\rho\over3(\rho+p)} \left( {k\over aH}
\right)^2 \Psi \,,
\end{equation}
where we have used to the constraint equation~(\ref{eq:Psi}) to
eliminate the comoving density perturbation, $\epsilon_m$. Note
that ${\cal R}$ and $-\zeta$ thus coincide in the limit $k\to0$.

Both ${\cal R}$ and $-\zeta$ are commonly used to characterise the
amplitude of adiabatic perturbations as both remain constant for
purely adiabatic perturbations on sufficiently large scales as a
direct consequence of local energy-momentum conservation~\cite{WMLL},
allowing one to relate the perturbation spectrum on large scales to
quantities at the Hubble scale crossing during inflation in the simplest
inflation models~\cite{BST,LL93}.

A dimensionless definition of the total entropy perturbation
(which is automatically gauge-invariant) is given by
\begin{equation}
\label{S_total}
{\cal S} = H \left( {\delta p \over \dot{p}} - {\delta\rho \over
\dot\rho} \right) \,,
\end{equation}
which can be extended to define a generalised entropy perturbation
between any two matter quantities
$x$ and $y$:
\begin{equation}
{\cal S}_{xy} = H \left( {\delta{x}\over \dot{x}} - {\delta{y}
\over \dot{y}} \right) \,.
\end{equation}
The total entropy perturbation in Eq.~(\ref{S_total}) for $N$ scalar
fields is given by 
\begin{equation}
\label{defSN} 
{\cal S} = \frac {
2\left( \dot{V} + 3H
\sum_J \dot\varphi_J^2 \right) \delta V 
 + 
2\dot{V}\sum_I \dot\varphi_I (
\dot{\delta\varphi}_I - \dot\varphi_I A )
} {3
\left(2\dot{V}+3H\sum_J \dot\varphi_J^2\right) \sum_I
\dot\varphi_I^2} \,,
\end{equation}
where the perturbation in the total potential energy is given by
$\delta V = \sum_I V_{\varphi_I}\delta\varphi_I$.

The change in ${\cal R}$ on large scales (i.e., neglecting spatial
gradient terms) can be directly related to the non-adiabatic part of
the pressure perturbation~\cite{GBW,WMLL,FB}
\begin{equation}
\dot{\cal R} \approx -3H {\dot{p}\over\dot\rho} {\cal S} \,.
\end{equation}
We will thus now consider the evolution of the adiabatic and entropy
perturbations in both one- and two-field models of inflation.

\subsection{Single field}

Perturbations in a single self-interacting scalar field obey the
gauge-dependent equation of motion
\begin{eqnarray}
\label{eq:varphi} && \ddot{\delta\varphi} + 3H\dot{\delta\varphi}
+ \left( \frac{k^2}{a^2}+V_{\varphi\varphi} \right) \delta\varphi
   \nonumber\\ &&~~{}=
-2V_{\varphi}A + \dot\varphi \left[ \dot{A} + 3\dot{\psi} +
\frac{k^2}{a^2} (a^2\dot{E}-aB) \right] \,,
\end{eqnarray}
subject to the energy and momentum constraint equations given in
Eqs.~(\ref{eq:densitycon}--\ref{eq:mtm}).

The scalar field perturbation in the spatially flat gauge has the
gauge-invariant definition, Eq.~(\ref{def:QI}),
\begin{equation}
Q_{\varphi} \equiv \delta\varphi +
{\dot\varphi\over H}\psi\,.
\end{equation}
For a single field this is directly related to the curvature
perturbation in the comoving gauge, where the momentum,
$\delta q=-\dot\varphi\delta\varphi$,
vanishes
\begin{equation}
\label{eq:zeta1}
{\cal R} = \psi + {H\over \dot\varphi} \delta\varphi = {H\over
\dot\varphi} Q_\varphi \,.
\end{equation}


It is not obvious that the intrinsic entropy perturbation for a single
scalar field, obtained from Eq.~(\ref{defSN}),
\begin{equation}
\label{eq:intS}
{\cal S}
 = {2V_\varphi \over 3\dot\varphi^2(3H\dot\varphi+2V_\varphi)}
 \left[ \dot\varphi \left( \dot{\delta\varphi} - \dot\varphi A \right) -
\ddot\varphi \delta\varphi\right] \,,
\end{equation}
should vanish on large scales. Because the scalar field obeys a
second-order equation of motion, its general solution contains two
arbitrary constants of integration, which can describe both
adiabatic and entropy perturbations.  However ${\cal S}$ for a
single scalar field is proportional to the comoving density
perturbation given in Eq.~(\ref{def:epsilonm}), and this in turn is
related to the metric perturbation, $\Psi$, via
Eq.~(\ref{eq:Psi}), so that~\cite{Bassett:1999mt}
\begin{equation}
\label{eq:S1}
{\cal S}
 = - {V_{\varphi} \over 6\pi G\dot\varphi^2[3H\dot\varphi+2V_\varphi]}
 \left( \frac{k^2}{a^2} \Psi \right) \,.
\end{equation}
In the absence of anisotropic stresses, $\Psi$ must be of order $A_Q$,
by Eq.~(\ref{eq:aniso}), and hence the non-adiabatic pressure becomes
small on large scales~\cite{SS,Bassett:1999mt,LR}. The amplitude of
the asymptotic solution for the scalar field at late times (and hence
large scales) during inflation thus determines the amplitude of an
adiabatic perturbation.

The change in the comoving curvature perturbation is given by
\begin{equation}\label{zetadot2}
\dot{\cal R} = {H\over \dot{H}} {k^2 \over a^2} \Psi
 \,,
\end{equation}
and hence the rate of change of the curvature perturbation, given
by $d\ln{\cal R}/d\ln a\sim (k/aH)^2$, becomes negligible on large
scales during single-field inflation.

\subsection{Two fields}

In this section we will consider two interacting scalar fields,
$\phi\equiv\varphi_1$ and $\chi\equiv\varphi_2$. The analysis
developed here should be straightforward to extend to include
additional scalar fields, but we do not expect to see any
qualitatively new features in this case, so for clarity we restrict
our discussion here to two fields.

In order to clarify the role of adiabatic and entropy
perturbations, their evolution and their inter-relation, we define
new adiabatic and entropy fields
by a rotation in field space. The ``adiabatic field'', $\sigma$,
represents the path length along the classical trajectory, such that
\begin{equation}
 \label{eq:sigma}
\dot\sigma = (\cos\theta) \dot\phi + (\sin\theta) \dot\chi \,,
\end{equation}
where
\begin{equation}
  \label{eq:cos sin}
\cos\theta = \frac{\dot{\phi}}{\sqrt{\dot{\phi}^2 +
\dot{\chi}^2}}, \quad \sin\theta =
\frac{\dot{\chi}}{\sqrt{\dot{\phi}^2 + \dot{\chi}^2}}\,.
\end{equation}
This definition, plus the original equations of motion for $\phi$ and
$\chi$, give
\begin{equation}
 \label{eq:sigma_dot_dot}
 \ddot{\sigma} + 3H\dot{\sigma} + V_\sigma = 0\,,
\end{equation}
where
\begin{eqnarray}
\label{eq:V_sigma}
V_\sigma &=& (\cos \theta) V_\phi + (\sin\theta) V_\chi\,.
\end{eqnarray}
As illustrated in Fig.~\ref{fig:decomposition}, $\delta \sigma$ is
the component of the two-field perturbation vector along the
direction of the background fields' evolution.
\begin{figure}[htbp]
  \begin{center}
\begin{picture}(0,0)%
\epsffile{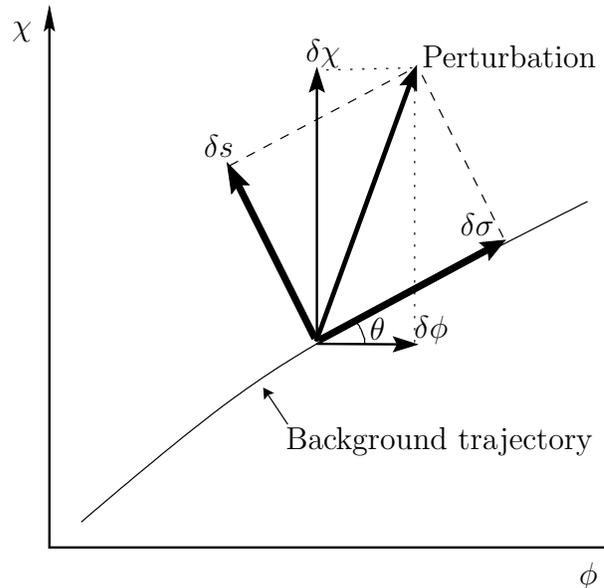}%
\end{picture}%
\setlength{\unitlength}{3947sp}%
\begingroup\makeatletter\ifx\SetFigFont\undefined%
\gdef\SetFigFont#1#2#3#4#5{%
  \reset@font\fontsize{#1}{#2pt}%
  \fontfamily{#3}\fontseries{#4}\fontshape{#5}%
  \selectfont}%
\fi\endgroup%
\begin{picture}(3897,3705)(1271,-3424)
\put(4085,-1171){\makebox(0,0)[lb]{\smash{\SetFigFont{12}{14.4}
{\rmdefault}{\mddefault}{\updefault}$\delta \sigma$}}}
\put(2990,-2517){\makebox(0,0)[lb]{\smash{\SetFigFont{12}{14.4}
{\rmdefault}{\mddefault}{\updefault}Background trajectory}}}
\put(3847,-113){\makebox(0,0)[lb]{\smash{\SetFigFont{12}{14.4}
{\rmdefault}{\mddefault}{\updefault}Perturbation}}}
\put(3116,-76){\makebox(0,0)[lb]{\smash{\SetFigFont{12}{14.4}
{\rmdefault}{\mddefault}{\updefault}$\delta \chi$}}}
\put(2476,-676){\makebox(0,0)[lb]{\smash{\SetFigFont{12}{14.4}
{\rmdefault}{\mddefault}{\updefault}$\delta s$}}}
\put(3798,-1813){\makebox(0,0)[lb]{\smash{\SetFigFont{12}{14.4}
{\rmdefault}{\mddefault}{\updefault}$\delta \phi$}}}
\put(3517,-1825){\makebox(0,0)[lb]{\smash{\SetFigFont{12}{14.4}
{\rmdefault}{\mddefault}{\updefault}$\theta$}}} \put(1271,
84){\makebox(0,0)[lb]{\smash{\SetFigFont{12}{14.4}{\rmdefault}
{\mddefault}{\updefault}$\chi$}}}
\put(4831,-3366){\makebox(0,0)[lb]{\smash{\SetFigFont{12}{14.4}
{\rmdefault}{\mddefault}{\updefault}$\phi$}}}
\end{picture}
\caption{
An illustration of the decomposition of an arbitrary perturbation
into an adiabatic ($\delta \sigma$) and entropy ($\delta s$)
component. The angle of the tangent to the background trajectory
is denoted by $\theta$. The usual perturbation decomposition,
along the $\phi$ and $\chi$ axes, is also shown.
}
    \label{fig:decomposition}
  \end{center}
\end{figure}
Conversely, fluctuations orthogonal to the
background classical trajectory represent non-adiabatic
perturbations, and we define the ``entropy field'', $s$, such that
\begin{gather}
\label{eq:s}
\delta s = (\cos\theta) \delta\chi - (\sin\theta) \delta\phi\,.
\end{gather}
From this definition, it follows that $s=$constant along the
classical trajectory, and hence entropy perturbations are
automatically gauge-invariant~\cite{StewartWalker}. Perturbations
in $\delta\sigma$, with $\delta s=0$, describe adiabatic
field perturbations, and this is why we refer to $\sigma$ as the
``adiabatic field''.

The total momentum of the two-field system, given by
Eq.~(\ref{eq:mtm}), is then
\begin{equation}
\delta q_{,i} = - \dot\phi\delta\phi_{,i} - \dot\chi\delta\chi_{,i}
= - \dot\sigma\delta\sigma_{,i} \,,
\end{equation}
and the comoving curvature perturbation in Eq.~(\ref{def:calR}) is
given by
\begin{eqnarray}
\label{eq:zeta}
{\cal R}
 &=& \psi + H\left( \frac{\dot\phi\delta\phi
              + \dot\chi\delta\chi}{\dot\phi^2 + \dot\chi^2}\right) 
\nonumber \,,\\
 &=& \psi + \frac{H}{\dot{\sigma}} \delta\sigma \,.
\end{eqnarray}
This expression, written in terms of the adiabatic field, $\sigma$,
is identical to that given in
Eq.~(\ref{eq:zeta1}) for a single field.

We can also write Eq.~(\ref{eq:zeta}) as
\begin{equation}
{\cal R}=(\cos^2\theta){\cal R}_\phi+(\sin^2\theta) {\cal
R}_\chi\,,
\end{equation}
where we define the comoving curvature perturbation for each of the
original fields as
\begin{equation}
{\cal R}_I \equiv \psi + {H\over\dot\varphi_I}\delta\varphi_I =
{H\over\dot\varphi_I}Q_I \,.
\end{equation}
However, even fields with no explicit interaction will in general have
non-zero intrinsic entropy perturbations on large scales in a
multi-field system due to their gravitational interaction, so that ${\cal
R}_{I}$ for each field is not conserved. Although the intrinsic
entropy perturbation for each field is still of the form given by
Eq.~(\ref{eq:intS}), it is no longer constrained by Eq.~(\ref{eq:Psi})
to vanish as $k\to0$.  This is in contrast to the case of
non-interacting perfect fluids, where it is possible to define a
constant curvature perturbation for each fluid on large
scales~\cite{WMLL}.

The comoving matter perturbation in Eq.~(\ref{def:epsilonm}) can
be written as
\begin{equation}
\label{epsilonm2} \epsilon_m = \dot\sigma \left(
\dot{\delta\sigma} - \dot\sigma A \right) -
\ddot\sigma\delta\sigma + 2 V_s \delta s \,,
\end{equation}
which acquires an additional term, compared with the single-field
case, due to the dependence of the potential upon $s$, where
\begin{equation}
V_s = (\cos\theta) V_\chi-(\sin\theta) V_\phi\,.
\end{equation}
The perturbed kinetic energy of $s$ has no contribution to
first-order as in the background solution $\dot{s}=0$, by
definition.

The total entropy perturbation, Eq.~(\ref{defSN}), for the two
fields can be written as
\begin{eqnarray}
&&{\cal S}
 = {2 \over 3\dot\sigma^2(3H\dot\sigma+2V_\sigma)}\times\nonumber\\
&&~~~{}\times \left\{ V_\sigma \left[ \dot\sigma \left(
\dot{\delta\sigma} - \dot\sigma A \right) - \ddot\sigma
\delta\sigma \right] + 3H\dot\sigma^2 \dot\theta\delta s \right\}
\,. \label{eq:S2}
\end{eqnarray}
Combining Eqs.~(\ref{eq:Psi}), (\ref{epsilonm2}) and~(\ref{eq:S2}), we
can write
\begin{equation}
{\cal S}
 = - {V_{\sigma} \over 6\pi G\dot\sigma^2[3H\dot\sigma+2V_\sigma]}
 \left( \frac{k^2}{a^2} \Psi \right)
- {2V_s \over3\dot\sigma^2} \delta s
\,.
\end{equation}
Comparing this with the single-field result given in
Eq.~(\ref{eq:S1}), we see that the entropy perturbation on large
scales is due solely to the relative entropy perturbation between
the two fields, described by the entropy field $\delta s$.

The change in the comoving curvature perturbation given by~\cite{GBW,FB}
\begin{equation}
\label{eq:zeta_dot_old}
  \dot{{\cal R}} = \frac{H}{\dot{H}}\frac{k^2}{a^2} \Psi
+{1\over2}H \left({\delta\phi\over\dot{\phi}}-{\delta
\chi\over\dot{\chi}}\right)
\frac{d}{dt}\left({\dot{\phi}^2-\dot{\chi}^2\over\dot{\phi}^2+
\dot{\chi}^2}\right) \, ,
\end{equation}
which can be expressed neatly in terms of the new variables:
\begin{equation}
  \label{eq:zeta_dot_new}
\dot{{\cal R}} = \frac{H}{\dot{H}}\frac{k^2}{a^2} \Psi +
\frac{2H}{\dot{\sigma}} \dot{\theta} \delta s \,,
\end{equation}
where
\begin{equation}
\dot{\theta} = -\frac{V_{s}}{\dot{\sigma}} \,.
\end{equation}
The new source term on the right-hand-side of this equation,
compared with the single-field case, Eq.~(\ref{zetadot2}), is
proportional to the relative entropy perturbation between the two
fields, $\delta s$.
Clearly, there can be significant changes to ${\cal R}$ on
large scales if the entropy perturbation is not
suppressed and if the background solution follows a curved
trajectory, i.e., $\dot\theta\neq0$, in field space~\cite{LR}.
This can then produce a change in the comoving curvature on
arbitrarily large scales (i.e., even in the limit
$k\to0$)~\cite{GBW,Bassett:1999mt}.

Equations of motion for the adiabatic and entropy field perturbations
can be derived from the perturbed scalar field
equations~(\ref{eq:perturbation}), to give
\begin{eqnarray} \label{eq:adiabatic1}
&& \delta \ddot{\sigma} + 3 H \delta \dot{\sigma} + \left(
\frac{k^2}{a^2} + V_{\sigma\sigma} - \dot{\theta}^2 \right)
 \delta \sigma \nonumber \\&&~~{}=
 -2V_{\sigma} A  + \dot{\sigma} \left[ \dot{A} +
3\dot\psi + \frac{k^2}{a^2} (a^2\dot{E}-aB) \right] \nonumber\\
&&~~~~~~{} +2(\dot\theta\delta s)^{\displaystyle{\cdot}} -
2{V_\sigma \over \dot\sigma} \dot\theta\delta s \,,
\end{eqnarray}
and
\begin{eqnarray}\label{eq:entropy1}
 &&\ddot{\delta s} + 3H\dot{\delta s} +
\left(\frac{k^2}{a^2}
  + V_{ss} - \dot{\theta}^2 \right) \delta s
\nonumber \\&&~~{}= -2{\dot\theta \over \dot\sigma} \left[
\dot\sigma ( \dot{\delta\sigma} - \dot\sigma A ) - \ddot\sigma
\delta \sigma \right] \,,
\end{eqnarray}
where 
\begin{eqnarray}
  \label{eq:V_sigma_sigma}
V_{\sigma\sigma}& = &(\sin^2\theta)V_{\chi\chi} +
(\sin2\theta)V_{\phi\chi} + (\cos^2\theta)V_{\phi\phi} \,,\\
  \label{eq:V_ss}
V_{ss} &=& (\sin^2\theta) V_{\phi\phi} - (\sin2\theta)V_{\phi\chi}
+ (\cos^2\theta)V_{\chi\chi}\,.
\end{eqnarray}
When $\dot\theta=0$, the adiabatic and entropy perturbations
decouple\footnote
{If we employ the slow-roll approximation for the background
fields, $\dot\phi\simeq -V_\phi/3H$ and $\dot\chi\simeq
-V_\chi/3H$, we obtain $\dot\theta\simeq0$. This reflects the fact
that the rate of change of $\theta$ is slow -- instantaneously it
moves in an approximately straight line in field space. But the
integrated change in $\theta$ cannot in general be neglected.
Even working within the slow-roll approximation, fields do not in
general follow a straight line trajectory in field space.}.
The equation of motion for $\delta\sigma$ then reduces to that for
a single scalar field in a perturbed FRW spacetime, as given in
Eq.~(\ref{eq:varphi}), while the equation for $\delta s$ is that
for a scalar field perturbation in an {\em unperturbed} FRW
spacetime.

The only source term on the right-hand-side 
in Eq.~(\ref{eq:entropy1})
for the entropy perturbation comes from the intrinsic entropy
perturbation in the $\sigma$-field. From Eqs.~(\ref{eq:Psi})
and~(\ref{epsilonm2}) we have
\begin{eqnarray}
\dot\sigma ( \dot{\delta\sigma} - \dot\sigma A ) - \ddot\sigma
\delta \sigma = 2\dot\sigma \dot\theta \delta s - {k^2\over 4\pi G
a^2} \Psi \,,\label{eq:int_entropy}
\end{eqnarray}
and hence we can rewrite the evolution
equation for the entropy perturbation as
\begin{equation}
\label{eq:entropy}
\ddot{\delta s} + 3H\dot{\delta s} + \left(\frac{k^2}{a^2}
  + V_{ss} + 3\dot{\theta}^2 \right) \delta s =
{\dot\theta\over\dot\sigma} {k^2 \over 2\pi G a^2} \Psi
\,.
\end{equation}
Note that this evolution equation is automatically gauge-invariant
and holds in any gauge. On large scales the inhomogeneous source
term becomes negligible, and we have a homogeneous second-order
equation of motion for the entropy perturbation, decoupled from
the adiabatic field and metric perturbations. If the initial
entropy perturbation is zero on large scales, it will remain so.

By contrast, we cannot neglect the metric back-reaction for the
adiabatic field fluctuations, or the source terms due to the
entropy perturbations. Working in the spatially flat gauge,
defining
\begin{equation}
Q_\sigma = \delta\sigma_Q = \delta\sigma + {\dot\sigma \over H}
\psi \,,
\end{equation}
and using
\begin{equation}
A_Q = 4\pi G {\dot\sigma \over H} Q_\sigma \,, \label{eq:A_Q}
\end{equation}
we can rewrite the equation of motion for the adiabatic field
perturbation as
\begin{eqnarray}
 && \ddot{Q}_\sigma + 3 H \dot{Q}_\sigma +
\left[ \frac{k^2}{a^2} + V_{\sigma\sigma} - \dot{\theta}^2 - {8\pi
G\over a^3} \left( {a^3\dot\sigma^2\over H}
\right)^{\displaystyle{\cdot}} \right] Q_\sigma \nonumber \\
&&~~~{}= 2(\dot\theta\delta s)^{\displaystyle{\cdot}} - 2\left(
{V_\sigma \over \dot\sigma} + {\dot{H}\over H} \right)
\dot\theta\delta s \,.\label{eq:adiabatic}
\end{eqnarray}
When $\dot\theta=0$, this reduces to the single-field equation of motion,
but for a curved trajectory in field
space, the entropy perturbation acts as an additional source term
in the equation of motion for the adiabatic field perturbation,
even on large scales.

In order for small-scale quantum fluctuations to produce
large-scale (super-Hubble) perturbations during inflation, a field
must be ``light'' (i.e., overdamped). The effective mass for the
entropy field in Eq.~(\ref{eq:entropy}) is
$\mu^2_s=V_{ss}+3\dot\theta^2$. For $\mu_s^2>{3\over2}H^2$, the
fluctuations remain in the vacuum state and fluctuations on large
scales are strongly suppressed. The existence of large-scale
entropy perturbations therefore requires
\begin{equation}\label{mus}
\mu^2_s\equiv V_{ss}+3\dot\theta^2 < {3\over2}H^2 \,.
\end{equation}

\section{Application to entropy/adiabatic correlations from inflation}

Equations~(\ref{eq:entropy}) and~(\ref{eq:adiabatic}) are the key
equations which govern the evolution of the adiabatic and entropy
perturbations in a two field system. Together with constraint
equations~(\ref{eq:int_entropy}) and~(\ref{eq:A_Q}) for the metric
perturbations, they form a closed set of equations. They allow one to
follow the effect on the adiabatic curvature perturbation due to the
presence of entropy perturbations, absent in the single field
model. This in turn will allow us to study the resulting correlations
between the spectra of adiabatic and entropy perturbations produced on
large-scales due to quantum fluctuations of the fields on small-scales
during inflation.

A useful approximation commonly made when studying field
perturbations during inflation, is to split the evolution of a
given mode into a sub-Hubble regime ($k>aH$), in which the Hubble
expansion is neglected, and a super-Hubble regime ($k<aH$), in
which gradient terms are dropped.

If we assume that both fields $\phi$ and $\chi$ are light (i.e.,
overdamped) during inflation, then we can take the field
fluctuations to be in their Minkowski vacuum state on sub-Hubble
scales. This gives their amplitudes at Hubble crossing ($k=aH$) as
\begin{equation}
  \label{eq:ics_old}
Q_I\big|_{k=aH} = \frac{H_k}{\sqrt{2k^3}}\,e_I(k)\,,
\end{equation}
where $I=\phi,\chi$, $H_k$ is the Hubble parameter when the mode
crosses the Hubble radius (i.e., $H_k=k/a$), and $e_\phi$ and
$e_\chi$ are independent Gaussian random variables satisfying
\begin{equation}
\label{eq:Gaussian} \langle e_I(k)\rangle = 0 \,, \quad \langle
e_I(k)e_J^*(k')\rangle = \delta_{IJ}\, \delta(k-k')\,,
\end{equation}
with the angled brackets denoting ensemble averages. It follows
from our definitions of the entropy and adiabatic perturbations in
Eqs.~(\ref{eq:sigma}) and~(\ref{eq:s}) that their distributions at
Hubble crossing have the same form:
\begin{equation}
  \label{eq:ics_new}
 Q_\sigma\big|_{k=aH} = \frac{H_k}{\sqrt{2k^3}}e_\sigma(k)\,,
 \quad
 \delta s\big|_{k=aH} = \frac{H_k}{\sqrt{2k^3}}e_s(k)\,,
\end{equation}
where $e_\sigma$ and $e_s$ are Gaussian random variables obeying
the same relations given in Eq.~(\ref{eq:Gaussian}), with
$I,J=\sigma,s$.

Super-Hubble modes are assumed to obey the equations of motion
given in Eqs.~(\ref{eq:adiabatic}) and~(\ref{eq:entropy}), which
we will write schematically as
\begin{eqnarray}
\label{eq:hatOsigma} \hat{O}^\sigma (Q_\sigma) &=& \hat{S}^\sigma
(\delta s)\,, \\ \hat{O}^s (\delta s) &=& 0\,,
\end{eqnarray}
where $\hat{O}^\sigma (Q_\sigma)$ and $\hat{O}^s (\delta s)$ are
obtained by setting $k=0$ on the left-hand side of
Eqs.~(\ref{eq:adiabatic}) and~(\ref{eq:entropy}) respectively, and
$\hat{S}^\sigma (\delta s)$ is given by the right-hand side of
Eq.~(\ref{eq:adiabatic}). As remarked before, there is no source
term for $\delta s$ appearing on the right-hand side of
Eq.~(\ref{eq:entropy}) once we neglect gradient terms. The general
super-Hubble solution can thus be written as
\begin{eqnarray}
Q_\sigma &=& A_+ f_+(t) + A_- f_-(t) + P(t) \,,\\ \delta s &=& B_+
g_+(t) + B_- g_-(t)\,,
\end{eqnarray}
where the real functions $f_{\pm}$ and $g_\pm$ are the
growing/decaying modes of the homogeneous equations,
$\hat{O}^\sigma(f_{\pm})=0$ and $\hat{O}^s(g_{\pm})=0$, and $P(t)$
is a particular integral of the full inhomogeneous
equation~(\ref{eq:hatOsigma}). 
Note that the growing-mode solution $f_+\propto \dot\sigma/H$.

Henceforth we shall consider only slow-roll inflation where the
evolution can be approximated by 
first-order equations [dropping $\ddot{\delta s}$ and
$\ddot{Q}_\sigma$ in Eqs.~(\ref{eq:entropy}) and~(\ref{eq:adiabatic})],
so that we have\footnote
{We note that in non-slow-roll scenarios the decaying modes
  may not be negligible on super-Hubble scales, which could 
  affect the correlations between adiabatic and entropy perturbations.}
\begin{eqnarray}
\label{SRsols}
Q_\sigma &\simeq& A f(t) + P(t) \,,\\
\delta s &\simeq& B g(t)\,.
\end{eqnarray}
We can, without loss of generality, take $f=1=g$ and $P=0$
when $k=aH$, so that the amplitudes of the growing modes at
Hubble-crossing are given by Eqs.~(\ref{eq:ics_new}) as
\begin{equation}
A(k) = \frac{H_k}{\sqrt{2k^3}}e_\sigma(k)\,,
 \quad
B(k) = \frac{H_k}{\sqrt{2k^3}}e_s(k)\,.
\end{equation}
From Eq.~(\ref{eq:hatOsigma}), we see that the amplitude of the
particular integral $P(t)$ at later times will be
correlated with the amplitude of the entropy perturbation, $B$,
and we can write $P(t)=B\tilde{P}(t)$, where $\tilde{P}(t)$ is a
real function
independent of the random variables $e_\sigma, e_s$.

In order to quantify the correlation, we define
\begin{equation}
\langle x(k) y^*(k') \rangle \equiv {2\pi^2\over k^3}\, {\cal
C}_{xy} \, \delta(k-k') \,.
\end{equation}
The adiabatic and entropy power spectra are given by
\begin{eqnarray}
{\cal P}_{Q_\sigma} \equiv
 {\cal C}_{Q_\sigma Q_\sigma}
 &\simeq& \left( {H_k\over2\pi} \right)^2
\left[ |f^2| + |\tilde{P}^2|\right] \,,\\ 
{\cal P}_{\delta s} \equiv 
 {\cal C}_{\delta s\delta s}
 &\simeq& \left( {H_k\over2\pi} \right)^2 |g^2| \,,
\end{eqnarray}
while the dimensionless cross-correlation is given by
\begin{equation}
\frac{{\cal C}_{Q_\sigma\delta s}} 
     {\sqrt{{\cal P}_{Q_\sigma}}\sqrt{{\cal P}_{\delta s}}}
 \simeq {g\tilde{P} \over \sqrt{g^2}\sqrt{|f^2|+|\tilde{P}^2|}} \,.
\end{equation}
Note that the adiabatic power spectrum at late times is always
enhanced if it is coupled to entropy perturbations [i.e., $P(t)\neq0$,
in Eq.~(\ref{SRsols})], as the entropy field fluctuations at
Hubble-crossing provide an uncorrelated extra source.

As an illustration, we consider the correlations in the adiabatic
and entropy perturbations at the start of the radiation era,
produced after double inflation, as studied in
Ref.~\cite{Langlois}. The double-inflation potential for two
non-interacting but massive scalar fields is:
\begin{equation}
  \label{eq:double potential}
V = \frac{1}{2}m_\phi^2 \phi^2 + \frac{1}{2}m_\chi^2 \chi^2\,.
\end{equation}
Following \cite{Polarski:1992dq}, it is possible to parametrise
the background scalar field trajectory in polar coordinates when
both fields are slow-rolling:
\begin{equation}
  \label{eq:di background}
\chi \simeq \sqrt{\frac{N}{2\pi G}}\sin\alpha\,, \quad \phi \simeq
\sqrt{\frac{N}{2\pi G}}\cos\alpha\,,
\end{equation}
where $N = -\ln(a/a_{\rm end})$ is the number of e-folds until the
end of inflation. The background trajectory can then be expressed
as:
\begin{equation}
  \label{eq:di background soln}
N \simeq N_0 \frac{(\sin\alpha)^{2/(R^2 - 1)}}
{(\cos\alpha)^{2R^2/(R^2 - 1)}}\,,
\end{equation}
where $R = {m_\chi}/{m_\phi}$. The scalar field position angle,
$\alpha$, can be related to the scalar field velocity angle,
$\theta$, which we used to define the adiabatic and entropy
perturbations:
\begin{equation}
  \label{eq:sin theta sin Theta}
\tan\theta \simeq -\frac{m_\chi^2}{3H\dot{\sigma}}
\sqrt{\frac{N}{2\pi G}}
\tan\alpha\,.
\end{equation}

The scalar field $\chi$ is assumed to decay into cold dark matter
while the scalar field $\phi$ decays into radiation. The
entropy/isocurvature at the start of the radiation-dominated era
is described by
\begin{equation}
  \label{eq:rad isoc}
S_{\rm rad} \equiv \frac{\delta \rho_c}{\rho_c} -
 {3\over4} \frac{\delta\rho_\gamma}{\rho_\gamma}\,.
\end{equation}
In Ref.~\cite{Langlois}, it is shown how the super-Hubble
perturbations in the radiation era can be determined in terms of
the perturbations during the inflationary era.  The fluctuations
in both $\phi$ and $\chi$ fields can contribute to both the
adiabatic and entropy perturbations. The adiabatic component comes
directly from the comoving curvature perturbation, ${\cal R}$, at
the end of inflation, and is given by
\begin{equation}
{\cal R}_{{\rm rad}} \simeq  -\sqrt{4\pi G} \sqrt{{N_k\over k^3}}
H_k\left[(\sin\alpha_k) e_\chi(k)  +(\cos\alpha_k)
e_\phi(k)\right]\,. \label{eq:Phi rad}
\end{equation}
The isocurvature perturbation at the start of the
radiation-dominated era is related to the entropy perturbation
between the two fields at the end of inflation~\cite{PolSta94}
\begin{equation}
S_{{\rm rad}} \simeq - {2\over3} m_\chi^2 {1\over H} \left(
{\delta\chi\over\dot\chi} - {\delta\phi\over\dot\phi} \right) \,,
\end{equation}
which yields
\begin{equation}
    \label{eq:entropy_entropy}
S_{{\rm rad}} \simeq -\sqrt{4\pi G}\sqrt{{N_k\over k^3}}H_k\left[ R^4
{\rm sec}\,\alpha_k + {\rm cosec}\,\alpha_k \right]e_s(k)\,,
\end{equation}
and
\begin{eqnarray}
\nonumber &&{\cal R}_{{\rm rad}} \simeq \sqrt{4\pi G}\sqrt{{N_k\over
k^3}} H_k
\frac{R^2\tan\alpha_k\sin\alpha_k}{\sqrt{R^2\tan^2\alpha_k +
1}}\times
\\&&~~{} \times \left\{\left[
\frac{1}{R^2\tan^2\alpha_k} + 1 \right]e_\sigma(k) +
\left[\frac{1-R^2}{R^2\tan\alpha_k}\right]e_s(k)  \right\}.
\label{eq:adiabatic_ent_adi}
\end{eqnarray}
The entropy perturbation during the radiation era only depends on
the entropy perturbation at Hubble-crossing during the
inflationary era, while the adiabatic perturbation during the
radiation era depends on both the adiabatic and entropy
perturbations at Hubble-crossing. This is consistent with
equations (\ref{eq:entropy}) and (\ref{eq:adiabatic1}), showing
that the entropy perturbation sources the adiabatic perturbation
on super-Hubble scales, but not vice versa.

As both equations (\ref{eq:entropy_entropy}) and
(\ref{eq:adiabatic_ent_adi}) depend on the random variable $e_s$,
the adiabatic and entropy perturbations will be correlated, and we
find
\begin{equation}
\frac{{\cal C}_{{\cal R}_{\rm rad}S_{\rm rad}}} 
     {\sqrt{{\cal P}_{{\cal R}_{\rm rad}}} \sqrt{{\cal P}_{S_{\rm rad}}}}
 \simeq \frac{(R^2-1)\sin
2\alpha_k}{2\sqrt{R^4\sin^2\alpha_k + \cos^2\alpha_k)}}\,.
\end{equation}
This correlation is investigated fully in \cite{Langlois} in terms
of the usual scalar field perturbation variables. An interesting
point that can easily be seen from
Eq.~(\ref{eq:adiabatic_ent_adi}) is that ${\cal R}_{{\rm rad}}$
will depend only on $e_\sigma$ if $R\equiv m_\chi/m_\phi=1$. Thus,
there will be no correlation if $R=1$. As can be seen from
Eq.~(\ref{eq:di background soln}), $\alpha$ will be constant for
$R=1$ and thus so will $\theta$; a straight-line background
trajectory will be obtained for $R=1$. This is consistent with
Eq.~(\ref{eq:adiabatic1}), where it can be seen that the entropy
component only sources the adiabatic component on large scales if
$\dot{\theta}\not=0$.

\section{Application to preheating after inflation}

In this section we use the entropy/adiabatic decomposition of the
perturbation equations to investigate the dynamics of super-Hubble
perturbations during a period of preheating at the end of
inflation. We consider three models, encompassed by the general
effective potential
\begin{equation}
  \label{eq:mod_pot}
V = \frac{1}{2} m^2\phi^2 + \frac{\lambda}{4}\phi^4 +
\frac{1}{2}g^2\phi^2\chi^2 + \tilde{g}^2 \phi^3\chi\,.
\end{equation}
The essence of preheating lies in the parametric amplification of
field perturbations due to the time-dependence of their effective
mass, e.g., $m_{\chi}^2\equiv V_{\chi\chi} = g^2\phi^2$. In the
simplest cases, the inflaton $\phi$ simply oscillates at the end of
inflation.

Preheating typically amplifies long-wavelength modes
preferentially. As discussed in~\cite{BKM1,Finelli:1999bu,Bassett:1999mt},
amplification of super-Hubble modes does not lead to a violation
of causality, due to the super-Hubble coherence of the inflaton
oscillations set up by the prior inflationary phase. If ${\cal R}$
is amplified on super-Hubble scales, this will alter the resulting
imprint on the anisotropies of the cosmic microwave background
(CMB), and break the simple link between CMB observations and
inflationary models.

We consider first the case where the inflaton is massive
($m\neq0$) and neglect its self-interaction ($\lambda=0$).
The traditional resonance parameter for the strength of preheating at
the end of inflation is
\begin{equation}
  \label{eq:q}
  q = \frac{g^2\phi_0^2}{4m^2}\,.
\end{equation}
In the massive case, where modes move through the resonance bands
of the Mathieu chart, and for inflation at high energies where the
expansion of the universe is very vigorous, $q$  needs to be much
larger than one if the parametric resonance is to be
efficient~\cite{Kofman:1997yn}.
It is possible to have large $q$ even for small coupling,
$g^2\ll1$, as $m\ll\phi_0\sim M_{\rm Pl}$.
We can write the effective mass of the $\chi$ during inflation as
\begin{equation}
  \label{eq:chi_mass}
\frac{m_\chi^2}{H^2} \approx \frac{3q}{\pi}\frac{M_{\rm Pl}^2}{\phi_0^2}\,,
\end{equation}
where $\phi_0$ is the initial value of $\phi$ at the beginning of
preheating. It then follows from Eq.~(\ref{eq:chi_mass}) that $\chi$
must be heavy during inflation for this simple potential if efficient
preheating is to be obtained.

Any change in the curvature perturbation ${\cal R}$ on very large
scales must be due to the presence of non-adiabatic perturbations.
In
\cite{Jedamzik:2000um,LLMW}, it was shown how, if $m_{\chi}^2 \gg
m_{\phi}^2$ during inflation
with $\lambda=0=\tilde{g}$, then the $\chi$ field and hence any
non-adiabatic perturbations on large scales are exponentially
suppressed during inflation, and no change to ${\cal R}$ occurs before
backreaction ends the resonance.

However, when $\tilde{g} \neq 0$, the $\chi$ field will have a nonzero
vacuum expectation value (vev) during inflation {\em even along the
valley of the potential}. 
In the slow-roll limit for $\phi$, this vev is determined by
$V_\chi=0$, which gives
\begin{equation}
  \label{eq:mod_traj}
  \chi \approx -\frac{\tilde{g}^2}{g^2}\phi\,.
\end{equation}
The $\tilde{g}$ coupling has the effect of rotating the valley of
the potential -- which the attractor trajectory approximately
follows -- from $\chi=0$, through an angle
\begin{equation}
  \label{eq:rotation}
  \theta \approx - \frac{\tilde{g}^2}{g^2}\,,
\end{equation}
where, to ensure that the chaotic inflation scenario is not drastically
altered, we assume~\cite{Bassett:2000ta} 
\begin{equation}
  \label{eq:small_g_tilde}
  \frac{\tilde{g}}{g} \ll 1\,.
\end{equation}

The effect of $\tilde{g}$ is to change the attractor for both $\chi$
and $\delta\chi$ during inflation, since the $\chi$ and $\delta\chi$
equations of motion gain inhomogeneous driving terms proportional to
$\tilde{g}^2\phi^3$.  
This does not necessarily imply that ${\cal R}$ will be amplified by
preheating at the end of inflation as purely adiabatic perturbations
along the slow-roll attractor now have a component along $\chi$ as
well as $\phi$.
In order to determine whether or not the evolution of the comoving
curvature perturbation, ${\cal R}$, on super-Hubble scales is
affected, we need to follow the evolution of the entropy field
perturbation\footnote{
{}From Eq.~(\ref{eq:zeta_dot_new}) we see that $\dot\theta\delta s$
must be non-zero to change ${\cal R}$ on large scales. 
Because $\dot\theta\approx0$, from Eq.~(\ref{eq:rotation}), the
entropy remains decoupled from the adiabatic perturbation during
slow-roll inflation in this model. But at the end of inflation, during
preheating, $\dot\theta\neq0$.
}, defined by Eq.~(\ref{eq:s}), which gives
\begin{equation}
\delta s \approx \delta\chi + \frac{\tilde{g}^2}{g^2} \delta\phi \,.
\end{equation}
In the limit $\tilde{g}/g\to0$ we recover $\delta s\to\delta\chi$.
Crucially, the evolution equation~(\ref{eq:entropy}) for the entropy
perturbation has {\em no} inhomogeneous terms in the long-wavelength
($k \rightarrow 0$) limit, even for $\tilde{g}\neq0$, and entropy
perturbations will only be non-negligible on super-Hubble scales if
the entropy field is light during inflation.

In the slow-roll limit and on large scales, the evolution
equation~(\ref{eq:entropy}) for the entropy perturbation has the
approximate solution~\cite{BD}
\begin{equation}
\delta s \propto a^{-3/2} \left(\frac{k}{aH}\right)^{-\nu}\,,
\end{equation}
where 
\begin{equation}
\nu^2 = \frac{9}{4} - \frac{\mu_s^2}{H^2}\,, \label{nus}
\end{equation}
and the effective mass of the entropy field, $\mu_{s}$ is defined in
Eq.~(\ref{mus}).
The power spectrum of entropy perturbations is
\begin{equation}
{\cal P}_{\delta s} \propto H^3 \left(\frac{k}{aH}\right)^{3-2
{\rm Re}(\nu)} \,. \label{powerspec2}
\end{equation}
The real part of $\nu$ vanishes for $\mu_s^2/H^2 > 9/4$,
leaving a steep $k^3$ blue spectrum, which is exponentially
suppressed with time.

Using Eqs.~(\ref{eq:V_ss}), (\ref{eq:mod_pot}), (\ref{eq:rotation}),
and (\ref{eq:small_g_tilde}), one finds that 
\begin{equation}
  \label{eq:mod_entropy_mass}
\frac{\mu_s^2}{H^2} \approx \left[1 - 4q\left({\tilde{g}\over
g}\right)^4\left({\phi\over\phi_0}\right)^2\right]^{-1}
\frac{3qM_{\rm Pl}^2}{\pi\phi_0^2}\,,
\end{equation}
$\mu_s^2/H^2$ has a local minimum for $\tilde{g}=0$. Thus the
additional $\tilde{g}$ term in Eq.~(\ref{eq:mod_pot}) serves to {\em
increase} the entropy mass relative to the Hubble parameter, and so
does not avoid the suppression of the entropy perturbation. The
$\tilde{g}$ term therefore does not significantly alter the spectral
index of the spectrum of entropy perturbations, which remains steep if
$q \gg 1$.  The strongly blue spectrum implies that non-linear
backreaction is dominated by small-scale modes, which go nonlinear
long before the cosmological modes, implying that resonance ends
before ${\cal R}$ changes~\cite{Kofman:1997yn,Jedamzik:2000um}.

We have also integrated the field equations numerically to avoid
relying on any slow-roll-type approximations.
To numerically evaluate the entropy perturbation, one could
simulate the original perturbation variables $\delta \phi$ and
$\delta \chi$, using Eq.~(\ref{eq:perturbation}), and then work
out $\delta s$ algebraically via Eq.~(\ref{eq:s}). However, this
approach is prone to numerical instability when the entropy
perturbation is suppressed.
To illustrate this, we take
$\tilde{g}=8\times 10^{-3}g$ and $q = 3.8\times10^5$ 
After about 60 e-folds of inflation, one can see analytically that
$\delta s \sim 10^{-40}$. Numerically, $\delta\chi\cos\theta \sim
\delta \phi\sin \theta\sim10^{-8}$ during inflation. So in order
to obtain a high enough accuracy to model the suppression of
$\delta s$, we require that $\delta\chi\cos\theta $ and $ \delta
\phi\sin \theta$ have to be simulated to a relative accuracy of
$\sim10^{-8}/10^{-40} = 10^{-32}$. This means approximately 32
significant figures are needed, which is beyond the capability of
standard numerical ordinary differential equation integration
routines.

If instead we use the new adiabatic and entropy field perturbations
and integrate Eqs.~(\ref{eq:entropy}) and (\ref{eq:adiabatic}), then
this numerical instability does {\em not} occur, since one no longer
needs to find the difference between two nearly equal
quantities.
Simulation results using these equations are compared
with the results using the old field perturbation
equations~(\ref{eq:perturbation}) in Fig.~\ref{fig:old_new}.
\begin{figure}[h]
\begin{picture}(0,0)%
\epsffile{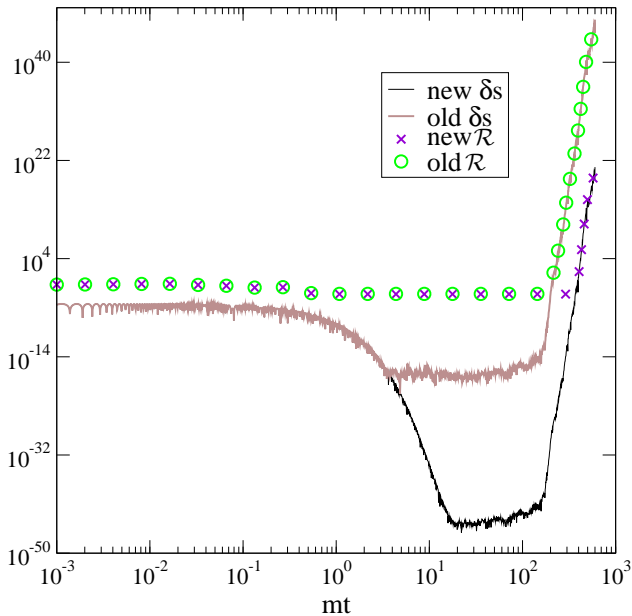}%
\end{picture}%
\setlength{\unitlength}{3947sp}%
\begingroup\makeatletter\ifx\SetFigFont\undefined%
\gdef\SetFigFont#1#2#3#4#5{%
  \reset@font\fontsize{#1}{#2pt}%
  \fontfamily{#3}\fontseries{#4}\fontshape{#5}%
  \selectfont}%
\fi\endgroup%
\begin{picture}(4107,4107)(3364,-6781)
\put(6376,-3741){\makebox(0,0)[lb]{\smash{\SetFigFont{9}{10.8}{\rmdefault}{\mddefault}{\updefault}$\cal R$}}}
\put(6321,-3896){\makebox(0,0)[lb]{\smash{\SetFigFont{9}{10.8}{\rmdefault}{\mddefault}{\updefault}$\cal R$}}}
\end{picture}
\caption{
 Numerical simulations of the entropy and comoving curvature
 perturbations during inflation and preheating, with $\lambda=0$,
 $g=2\times10^{-3}$, $\tilde{g}=8\times10^{-3}g$ and $m=10^{-6}M_{\rm
   pl}$.  The
 `new' prefix indicates that the field perturbations were evaluated by
 numerically integrating Eqs.~(\ref{eq:entropy}), and
 (\ref{eq:adiabatic}), while the `old' prefix indicates that the
 perturbations were evaluated by integrating the original field
 equations~(\ref{eq:perturbation}).  We have not included any
 higher-order corrections such as backreaction from small-scale
 perturbations which would shut down the resonant amplification of
 $\delta s$ at some point.
 }
     \label{fig:old_new}
 \end{figure}
 The
simulations show that the growth in $\cal R$ is driven by $\delta s$,
in concordance with Eq.~(\ref{eq:zeta_dot_new}). As can be seen, the
numerical result using the field perturbation equations fails to track
the exponential decay of the entropy during inflation and thus
underestimates the delay in the growth of $\cal R$.

In practice, we find a similar instability if we
try to construct the gauge-invariant metric perturbation, $\Psi$,
required in Eq.~(\ref{eq:entropy}) in terms of the constraint
Eq.~(\ref{eq:int_entropy}). This includes the intrinsic entropy
perturbation in the $\sigma$ field, which does become small at late
times/large scales, but results from the diminishing difference between
finite terms. It is more stable numerically to follow the value of
$\Psi$ at late times using the evolution equation
\begin{equation}
\dot\Psi + \left( H - {\dot{H}\over H} \right) \Psi = 4\pi G\dot\sigma
Q_\sigma \,,
\end{equation}
which can be obtained from the definition of $\Psi$ given in
Eq.~(\ref{eq:Psi}) and the metric constraint
equations~(\ref{eq:mtmcon}) and~(\ref{eq:aniso}).

Note that the adiabatic/entropy decomposition becomes
ill-defined if $\dot\sigma=0$, i.e. both fields stop rolling, and this can
cause numerical instability during preheating if the trajectory is
confined to a narrow valley. 
This can occur, for instance, when $\tilde{g}=0$ and only the $\phi$
field oscillates.
The original field perturbations $\delta\phi$ and $\delta\chi$ remain
well-defined, although the comoving curvature perturbation ${\cal R}$,
defined in  Eq.~(\ref{eq:zeta}) becomes singular when
$\dot\sigma=0$~\cite{HK}. 
This does not happen for the simulation results shown in
Fig.~\ref{fig:old_new} with $\tilde{g}\neq0$ where the fields
oscillate in a two-dimensional potential well.

The massive inflaton potential ($m\neq0$) safeguards the conservation
of ${\cal R}$ by a bootstrap effect: if preheating is strong, $q \gg
1$, then the entropy perturbation is heavy during inflation; on the
other hand, if the entropy is light during inflation, then $q \leq 1$
and preheating is very weak. This is not altered by a rotation of the
trajectory in field space ($\tilde{g}\neq0$) as can be most quickly
seen by noting, from Eqs.~(\ref{eq:V_sigma_sigma})
and~(\ref{eq:V_ss}), that
\begin{equation}
\label{sumrule}
V_{\sigma\sigma} + V_{ss} = V_{\phi\phi} + V_{\chi\chi} \,.
\end{equation}
Thus if the $\chi$ field is very massive ($V_{\chi\chi}\gg H^2$), we
must have $V_{\sigma\sigma} + V_{ss}\gg H^2$. For slow-roll inflation
we require $V_{\sigma\sigma}\ll H^2$ and hence $V_{ss}\gg H^2$.

This situation does not hold if the entropy mass during inflation is
not linked to the entropy mass during
preheating~\cite{Bassett:2000ta}, or in massless ($m=0$)
self-interacting ($\lambda\neq0$) inflation models~\cite{BV,FB,ZBS}.
This latter class of models are almost conformally invariant, allowing
analytical results from Floquet theory to be applied.  The Floquet
index, $\mu_k$, which determines the rate of exponential growth, can
reach its maximum as $k/aH \to 0$, when $g^2/\lambda = 2 n^2$ for
integer $n$, thereby implying maximum growth for the
longest-wavelength perturbations.  
Assuming slow-roll inflation driven by $V\approx \lambda\phi^4/4$, we
see from Eq.~(\ref{sumrule}) that $V_{\sigma\sigma} +
V_{ss}>V_{\chi\chi}=g^2\phi^2$ and thus that the entropy field is
massive ($V_{ss}>9H^2/4$) whenever
\begin{equation}
\frac{g^2}{\lambda} > 8\pi \frac{\phi^2}{M_{\rm Pl}^2} \,.
\end{equation}
However, we can have resonance at large scales for $n=1$ and
$g^2/\lambda = 2$, when the entropy field need not be heavy during
inflation and no exponential suppression takes place, so that the
subsequent growth of ${\cal R}$ is explosive~\cite{BV}. The growth of
${\cal R}$ occurs before backreaction can shut off the resonant growth
of the entropy perturbations $\delta s$~\cite{BV,FB,TBV,ZBS}.
Although the region of parameter space around $g^2/\lambda = 2$ is
thus ruled out, the same does not hold for $g^2/\lambda \gg 1$, since
the entropy field is then heavy during inflation and $\delta s$ is
again suppressed.  

\section{Conclusions}

We have introduced a new formalism in which to follow the evolution of
adiabatic and entropy perturbations during inflation with multiple
scalar fields.  We decompose arbitrary field perturbations into a
component parallel to the background solution in field space, termed
the {\em adiabatic\/} perturbation, and a component orthogonal to the
trajectory, termed the {\em entropy\/} perturbation. We have rederived
the field equations in terms of these rotated fields in
Eqs.~(\ref{eq:entropy}) and~(\ref{eq:adiabatic}). These show that the
adiabatic perturbation on large scales can be driven by the entropy
perturbation, while the entropy perturbation itself obeys a
homogeneous second-order equation on super-Hubble scales.  There can
only be significant change in the large-scale comoving curvature
perturbation if there is a non-negligible entropy perturbation, {\em and}
if the background trajectory in field space is curved.

Our formalism can be applied to evaluate the correlation between the
adiabatic and entropy perturbations at the end of inflation.  As an
example we considered the example of two non-interacting fields in
double inflation, calculating the cross-correlation between the
adiabatic and entropy perturbations.

The effect of preheating on the large-scale curvature perturbation can
also be addressed within our formalism.  The mass of the entropy field
during inflation is a crucial quantity. If the entropy field is heavy,
then any fluctuations on large scales are suppressed to negligible
values at the beginning of preheating. This squeezing of the entropy
perturbation is most accurately modelled numerically using our
evolution equation for the entropy perturbation. If it is estimated
from the usual field equations, it may contain large numerical errors
when there is a non-trivial background trajectory in field space.

\section*{Acknowledgements}
We thank Jai-chan Hwang, David Langlois, Andrew Liddle, David Lyth,
Karim Malik and Shinji 
Tsujikawa for useful comments and discussions. DW is supported by the Royal 
Society.

{\em Note added:} After completing this work we became aware of
related work by Hwang and Noh~\cite{Hwang+Noh} who also study entropy
perturbations in multiple field inflation. They find that the
adiabatic and entropy modes decouple on super-horizon scales when
the effect of curvature of the trajectory in field
space is neglected, but we have shown that this cannot in general be
assumed, even in models of slow-roll inflation.

\end{document}